\documentclass[twocolumn,prl,superscriptaddress,notitlepage]{revtex4-1}

\usepackage{graphicx}
\usepackage{comment}
\usepackage{amsmath,amssymb,amsthm} 
\usepackage{verbatim}
\usepackage{float}

\setcitestyle{super}

\usepackage[sort&compress]{natbib}

\begin{document}

\title{The Rosenfeld-Tarazona expression for liquids' specific heat: A numerical investigation of eighteen systems}
\author{Trond S. Ingebrigtsen}
\email{trond@ruc.dk}
\affiliation{DNRF Centre ``Glass and Time'', IMFUFA, Department of Sciences, Roskilde University, Postbox 260, DK-4000 Roskilde, Denmark}
\author{Arno A. Veldhorst}
\affiliation{DNRF Centre ``Glass and Time'', IMFUFA, Department of Sciences, Roskilde University, Postbox 260, DK-4000 Roskilde, Denmark}
\author{Thomas B. Schr{\o}der}
\affiliation{DNRF Centre ``Glass and Time'', IMFUFA, Department of Sciences, Roskilde University, Postbox 260, DK-4000 Roskilde, Denmark}
\author{Jeppe C. Dyre}
\affiliation{DNRF Centre ``Glass and Time'', IMFUFA, Department of Sciences, Roskilde University, Postbox 260, DK-4000 Roskilde, Denmark}

\date{\today}

\begin{abstract}
  We investigate the accuracy of the expression of Rosenfeld and Tarazona (RT) for the excess 
  isochoric heat capacity, $C_{V}^{\mbox{ex}} \propto T^{-2/5}$, for eighteen model liquids. Previous investigations have reported no unifying 
  features of breakdown for the RT expression. Here liquids with different 
  stoichiometric composition, molecular topology, chemical interactions, degree of undercooling, and environment are investigated. We find that the RT expression is a 
  better approximation for liquids with strong correlations between equilibrium fluctuations of virial and 
  potential energy, i.e., Roskilde simple liquids [Ingebrigtsen \textit{et al.}, Phys. Rev. X \textbf{2}, 011011 (2012)]. This 
  observation holds even for molecular liquids under severe nanoscale confinement, the physics of which is completely different from the original RT bulk 
  hard-sphere fluid arguments. The density dependence of the specific heat is predicted from the isomorph theory for Roskilde simple liquids, which 
  in combination with the RT expression provides a complete description of the specific heat's density and temperature dependence.
\end{abstract}

\maketitle

Fundamental theories for the pressure (or density) and temperature dependence of thermodynamic quantities have gained renewed attention in the last decade. These 
theories can serve as a valuable input to equations of state\cite{spera2009,paper5}, but also as input to scaling strategies which relate key 
dimensionless transport coefficients to thermodynamic quantities, such as the excess 
entropy\cite{rosenfeld1,dzugutov1996} (with respect to an ideal gas) or the excess isochoric heat capacity\cite{ingebrigtsen2013}. Predicting dynamical quantities 
from first principles is a challenging task. One such theory is 
mode-coupling theory\cite{gotze2008} (MCT) which relates the dynamic density correlations 
of a fluid to its static structure. However, MCT becomes less reliable as the characterizing features of 
supercooled liquids become pronounced\cite{debenedetti2001,ka1,ka2}. Extensions of MCT to more complex environments, such as nanoscale confinement or homogenous shear flow, have 
been slow to emerge\cite{krakoviack2005,biroli2006,lang2010,lang2012,suzuki2013}. Recently, however, other promising MCT-type-theories have started to appear for supercooled liquids\cite{schweizer2005}.

Alternatively, theories which relate dynamics to thermodynamics such as that of Adam and Gibbs\cite{adam1965,dyre2009} and Rosenfeld's excess 
entropy scaling\cite{rosenfeld1,rosenfeld2}, relate to the dynamics also in the highly supercooled liquid regime\cite{mittal2006,sengupta2012}. Excess entropy 
scaling strategies, as proposed by Rosenfeld, have proven succesful in predicting the dynamics of not only
single-component atomic fluids\cite{rosenfeld1,dzugutov1996}, but also binary mixtures\cite{dzugutov1996,mittal2006,sengupta2012,singh2012}, ionic substances\cite{singh2011,jabes2012}, small 
molecules\cite{truskdumbbellbulk,truskwater,truskhydrocarbon,abramson2008,abramson2009}, and polymers\cite{excessdensityscaling}. In fact, excess entropy scaling strategies 
have been used as a reliable predictor even for the perplexing dynamics of nanoconfined liquids\cite{hsconfined,LJconfinedsmoothwalls,
  connectiondynamicsprofile,dumbbellconfinedroughwalls,goel2009,borah2012,ingebrigtsen2013} which exhibit stratification and position-dependent 
relaxation processes; the applicability of excess entropy scaling strategies to these situations emphasizes the usefulness of the thermodynamic approach to dynamics. 

To fully harness the power of predicting dynamics from thermodynamics, however, it is imperative to develop reliable theories for the 
pressure (or density) and temperature dependence of thermodynamic quantities. Rosenfeld and Tarazona\cite{RT} (RT) argued 
for a mathematically simple expression for the density and temperature dependence of the potential energy $U$ for fluids. Their arguments are based on
thermodynamic perturbation theory, using the fundamental-measure reference functional for hard-spheres in combination 
with an expansion of the free energy around the $\eta = 1$ packing fraction. The arguments are involved and not easy to follow, but 
their expressions have found widespread application (Refs. \onlinecite{RT,leonardo2000,gebre2005,may2012,coluzzi1999,sciortino1999,
  coluzzi2000,fernandez2006,foffi2008,frankcasperbonds,paper5,shi2011,allsopp2012,saika2001,agarwal2007,spera2009,jabes2010,
  scala2000,palomar2005,sharma2008,agarwal2010,truskwater,agarwal2011,mossa2002,truskhydrocarbon,angelani2002}).

From the potential energy one gains access to thermodynamic quantities such as the excess isochoric heat capacity $C_{V}^{\mbox{ex}} = (\partial U / \partial T)_{V}$  and the
excess entropy $S_{\mbox{ex}}$ via $C_{V}^{\mbox{ex}} = T (\partial S_{\mbox{ex}} / \partial T)_{V}$. Both of these quantities enter the aforementioned strategies.
For a long time only few studies focused on the heat capacity\cite{wallace1960}. Recently, however,  
the heat capacity of liquids has started to receive more attention\cite{davatolhagh2005,trachenko2008,trachenko2011,bolmatov2012,tatsumi2012}.
The RT expressions for the potential energy and excess isochoric heat capacity read

\begin{align}
  U(\rho, T) & = \alpha(\rho)T^{3/5} + \beta(\rho),\label{RTeq} \\
  C_{V}^{\mbox{ex}}(\rho, T) & = 3/5\alpha(\rho)T^{-2/5},\label{RT2eq}
\end{align}
where $\alpha(\rho)$ and $\beta(\rho)$ are extensive functions of density $\rho$ that relate to the specific system\cite{RT}.

Several numerical investigations have tested the applicability of the RT expressions for various model liquids. 
These liquids span from simple atomic model fluids to liquids showing a wide range of structural, dynamical, and thermodynamical anomalies in their phase diagram. 
More specifically, the RT expressions have been 
investigated for single-component atomic fluids\cite{RT,leonardo2000,gebre2005,may2012}, 
binary mixtures\cite{coluzzi1999,sciortino1999,coluzzi2000,fernandez2006,foffi2008,frankcasperbonds,paper5,shi2011,allsopp2012}, 
ionic substances\cite{saika2001,agarwal2007,spera2009,jabes2010}, 
hydrogen-bonding liquids\cite{scala2000,palomar2005,sharma2008,agarwal2010,truskwater,agarwal2011}, 
small molecules\cite{mossa2002,truskhydrocarbon}, and sheared liquids\cite{angelani2002}. 
These investigations showed that the RT expressions give a good approximation for a range of systems, but are less accurate
when applied to systems known to not have strong virial/potential energy correlations\cite{paper1}, 
such as the Dzugutov liquid and Gaussian core model, as well as for $\mathrm{SiO_{2}}$ and $\mathrm{BeF_{2}}$ in their anomalous regions. For SPC/E water 
different results for the applicability have been reported\cite{scala2000,truskwater,agarwal2010}. 

The purpose of this paper is to provide insight 
into the conditions under which RT applies  by investigating 18 different model systems possessing  different stoichiometric composition, molecular topology, chemical interactions, degree 
of undercooling, and environment. We use GPU-optimized\cite{rumd} \textit{NVT} molecular dynamics computer 
simulations\cite{nose,hoover,nvttoxvaerd,toxconstraintnve} (in total over 40000 GPU hours) to 
calculate the potential energy and excess isochoric heat capacity along a single isochore for 
each of these 18 model systems (for the single-component Lennard-Jones (SCLJ) liquid we also vary the density). Here and 
henceforth quantities are reported in dimensionless units, e.g., by setting $\sigma = 1$, $\epsilon = 1$, etc.
The heat capacity is calculated via Einstein's fluctuation formula $C_{V}^{\mbox{ex}} = \langle (\Delta U)^{2}\rangle / k_{B}T^{2}$ (within 
statistical error, the numerical derivative of the potential energy agrees with the value calculated from the fluctuations). Previous investigations have often 
calculated the excess entropy, but this is computationally very demanding and not necessary to test Eqs. (\ref{RTeq}) and (\ref{RT2eq}). Table \ref{RTTAB} presents 
the investigated model systems, which range from simple atomic fluids to molecules under severe nanoscale confinement. The densities represent typical liquid-state densities.

\begin{figure}[H]
  \centering
  \includegraphics[width=75mm]{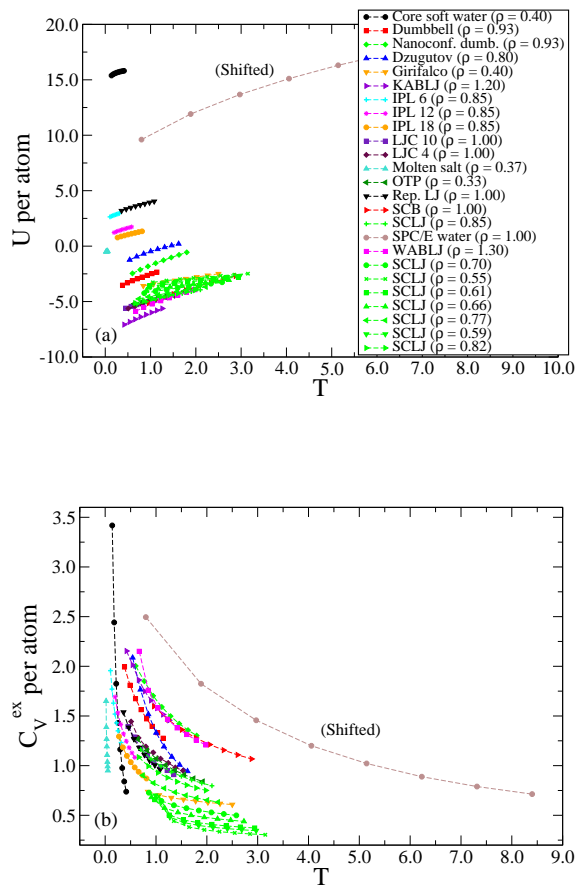}
  \caption{The isochores studied for each of the $18$ different model systems. (a) The potential energy $U$ per atom as a function of temperature (the 
    values for SPC/E water have been shifted for clarity with -3 in the $x$-direction and 33 in the $y$-direction).
    (b) The excess isochoric heat capacity $C_{V}^{\mbox{ex}} = \langle (\Delta U)^{2} \rangle / k_{B}T^{2}$ per atom as a function of temperature (the 
    values for SPC/E water have been shifted for clarity with -3 in the $x$-direction). For all liquids $C_{V}^{\mbox{ex}}$ decreases with increasing temperature.}
    
  \label{RT1}
\end{figure}

\begin{table}[H]
  \scriptsize
  \centering
  \begin{tabular}
    {|| p{3.2cm} || p{0.6cm} | p{0.8cm}| p{1.1cm} | p{1.1cm} | p{0.6cm} || }
    \hline
    \hline 
    \textbf{System} & $\rho$ & $T_{min}$ & $D_{U}$ & $D_{C_{V}^{\mbox{ex}}}$ & $R$ \\
    \hline 
    Core-soft water\cite{coresoft}           & 0.40 & 0.138  & 0.974     & 0.473     & 0.10 \\
    Dumbbell\cite{moleculeshidden}           & 0.93 & 0.380  & $>$ 0.999 & 0.999     & 0.96 \\
    Nanoconfined dumbbell\cite{ingebrigtsen2013} & 0.93 & 0.600  & $>$ 0.999 & 0.998     & 0.91 \\
    Dzugutov\cite{dzugutov}                  & 0.80 & 0.540  & 0.997     & 0.786     & 0.71 \\
    Girifalco\cite{girifaclo1992}            & 0.40 & 0.840  & 0.999     & -0.664    & 0.91 \\
    KABLJ\cite{ka1}                          & 1.20 & 0.420  & $>$ 0.999 & 0.984     & 0.93 \\
    \hline
    IPL 6                                    & 0.85 & 0.104  & $>$ 0.999 & 0.997     & 1.00 \\
    IPL 12                                   & 0.85 & 0.195  & $>$ 0.999 & $>$ 0.999 & 1.00 \\
    IPL 18                                   & 0.85 & 0.271  & $>$ 0.999 & 0.988     & 1.00 \\
    LJC 10\cite{ljc}                         & 1.00 & 0.450  & $>$ 0.999 & 0.998     & 0.86 \\
    LJC 4                                    & 1.00 & 0.510  & $>$ 0.999 & 0.991     & 0.90 \\
    Molten salt\cite{moltensalt}             & 0.37 & 0.018  & $>$ 0.999 & 0.952     & 0.15 \\
    OTP\cite{otp1}                           & 0.33 & 0.640  & $>$ 0.999 & 0.995     & 0.91 \\
    Repulsive LJ\cite{thermoscl}             & 1.00 & 0.360  & $>$ 0.999 & 0.995     & 1.00 \\
    SCB\cite{buckingham1}                    & 1.00 & 0.960  & $>$ 0.999 & 0.991     & 0.99 \\
    \hline
    SCLJ                                     & 0.85 & 0.700  & $>$ 0.999 & 0.974     & 0.96 \\
    SCLJ                                     & 0.82 & 0.660  & $> 0.999$ & 0.962     & 0.94 \\
    SCLJ                                     & 0.77 & 0.740  & $>$ 0.999 & 0.940     & 0.90 \\
    SCLJ                                     & 0.70 & 0.860  & $>$ 0.999 & 0.954     & 0.82 \\
    SCLJ                                     & 0.66 & 0.910  & $>$ 0.999 & 0.959     & 0.75 \\
    SCLJ                                     & 0.61 & 0.980  & $>$ 0.999 & 0.859     & 0.64 \\
    SCLJ                                     & 0.59 & 0.990  & $>$ 0.999 & 0.729     & 0.56 \\
    SCLJ                                     & 0.55 & 1.050  & $>$ 0.999 & 0.644     & 0.51 \\
    SPC/E water\cite{spce}                   & 1.00 & 3.800  & 0.987     & 0.558     & 0.07 \\
    WABLJ\cite{wahnstrom1991}                & 1.30 & 0.670  & $>$ 0.999 & 0.911     & 0.98 \\
    \hline 
    \hline 
  \end{tabular}
  \caption{Model systems investigated. $D_{U}$ 
    and $D_{C_{V}^{\mbox{ex}}}$ are the coefficient of determination (Eq. (\ref{detcoef})) for the potential energy and excess isochoric heat capacity, respectively, for the 
    isochore of density $\rho$. The virial/potential energy correlation coefficient $R$ is given for the lowest temperature state point $T_{min}$. 
    The abbreviations used are: Kob-Andersen binary Lennard-Jones mixture (KABLJ); inverse power-law fluid with exponent $n$ (IPL $n$); LJ polymer chain of length $n$ (LJC $n$); 
    Lewis-Wahnstr\"om $o$-terphenyl (OTP); single-component Buckingham liquid (SCB); single-component LJ liquid (SCLJ); Wahnstr\"om binary LJ mixture (WABLJ). The ''Nanoconfined dumbbell'' 
    is confined to a (smooth) slit-pore of width $H = 8.13$, corresponding to roughly 16 molecular lengths. }
  \label{RTTAB}
\end{table}
Figures \ref{RT1}(a) and (b) show, respectively, the potential energy and excess isochoric heat capacity 
at constant density as a function of temperature for all investigated systems. In all cases, the 
excess isochoric heat capacity decreases with increasing temperature. \newline
\begin{figure}[H]
  \centering
  \includegraphics[width=75mm]{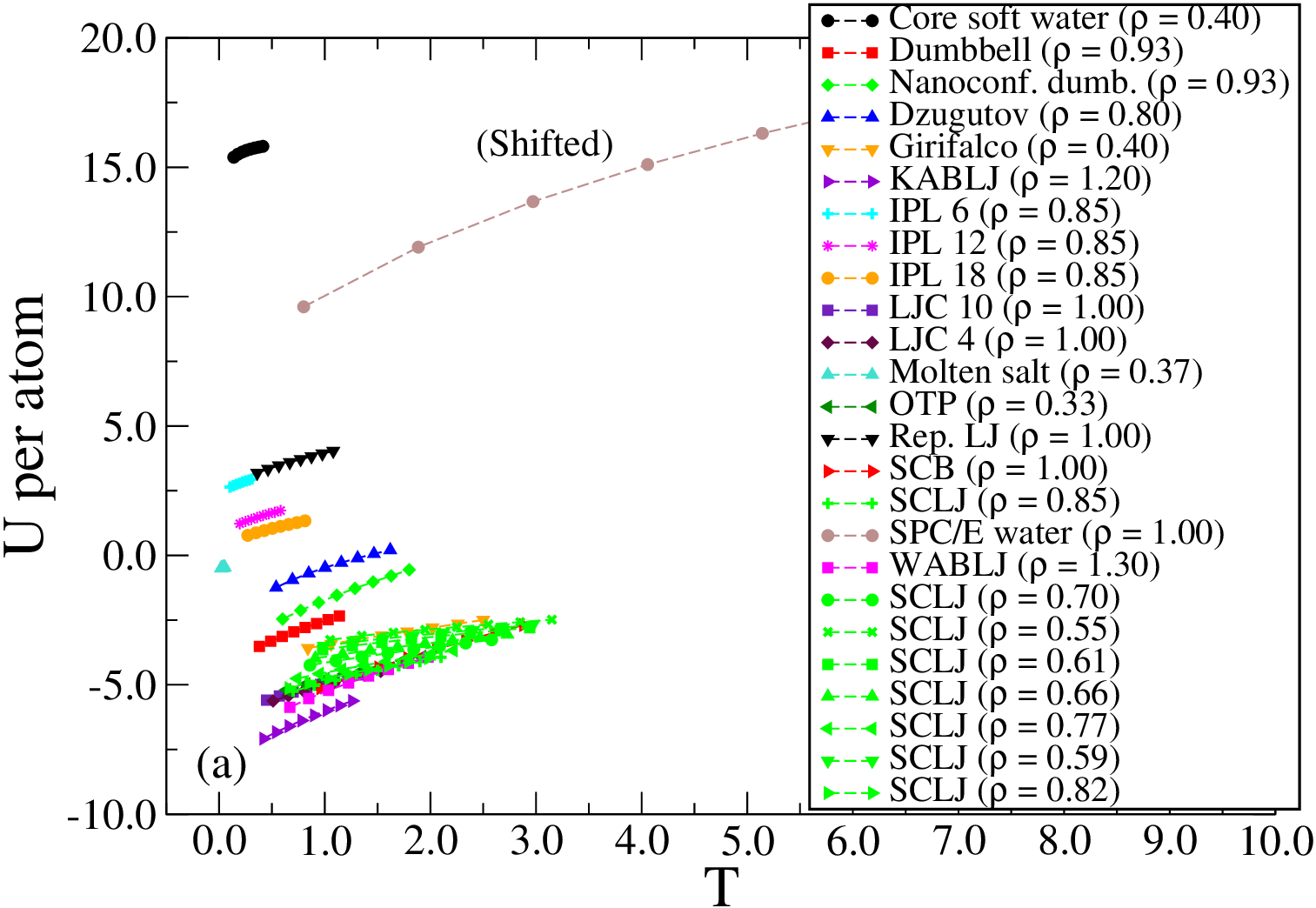}
\end{figure}
\hspace{5pt}
The data points of Fig. \ref{RT1} were generated by the following procedure.

\begin{figure}[H]
  \includegraphics[width=75mm]{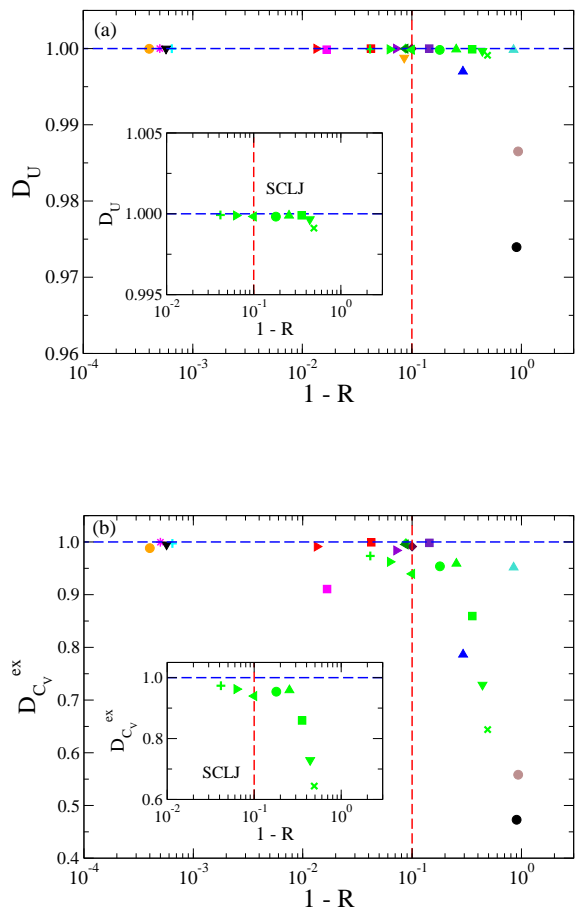}
  \centering
  \caption{The coefficient of determination $D$ (Eq. (\ref{detcoef})) for $U$ and $C_{V}^{\mbox{ex}}$ as a function of $1-R$ for 
    the $18$ different model systems. The insets show $D$ for the SCLJ liquid (see Table \ref{RTTAB}). The definition 
    of the symbols is given in Fig. \ref{RT1}. (a) $D_{U}$. (b) $D_{C_{V}^{\mbox{ex}}}$. The Girifalco system gives a negative value for $D_{C_{V}^{\mbox{ex}}}$ and 
  has for clarity of presentation been left out (see Table \ref{RTTAB}). For both the potential energy and the excess isochoric heat capacity, RT is seen 
  to deteriorate as $R$ decreases below 0.90 (to the right of the red line); in particular see insets for the SCLJ liquid. }
  \label{RT2}
\end{figure}

\begin{enumerate} 
\item First the system is cooled at constant density until one of the following happens: a) The system 
  crystallizes; b) The pressure becomes negative; c) The relaxation time is of the order $10^{5}$ time units. 
  This happens at the temperature $T_{min}$. The system is then equilibrated at $T$ = $T_{min}$; in the case of 
  crystallization or negative pressure, the temperature is increased slightly (and this new temperature defines $T_{min}$).
\item Next, the temperature is increased from $T_{min}$ up to $T_{max} = 3T_{min}$, probing state points along the isochore 
  with a spacing of $\Delta T = (T_{max} - T_{min})/7$. A total of eight equilibrium state points are hereby generated for each isochore.
\end{enumerate}
Turning now to the RT expressions, we show in Figs. \ref{RT2}(a) and (b)
the coefficient of determination\cite{coef} $D$ for the potential energy and excess isochoric heat capacity as a 
function of $1 - R$ (see below). For a generic quantity $X$, the coefficient of determination $D_{X}$ is defined by 

\begin{align}
  D_{X} & = 1 - \frac{\sum_{i=1}^{N}\big(X_{i} - f({X_{i})}\big)^{2}}{\sum_{i=1}^{N}(X_{i} - \langle X \rangle)^{2}},\label{detcoef}
\end{align}
where $f({X}_{i})$ is a function that provides the model values, and the average $\langle X \rangle$ is taken over a set of data 
points $\textbf{X} = \{X_{1}, ..., X_{N} \}$. In our case $f({X}_{i})$ is given by fits to 
the data points in \textbf{X} using, respectively, $U = A_{0}\,T^{3/5} + A_{1}$, and $C_{V}^{\mbox{ex}} = 3/5A_{2}\,T^{-2/5}$, where 
$A_{0}$, $A_{1}$, and $A_{2}$ are constants. $D_{X}$ measures the proportion of variability in a data set that is accounted for by the 
statistical model\cite{coef}; $D_{X}$ = 1 implies perfect account of the variability.  

The virial/potential energy correlation coefficient $R$ is defined\cite{paper1} via
\begin{equation}
R = \frac{\langle \Delta W \Delta U \rangle}{\sqrt{\langle (\Delta W)^{2} \rangle}\sqrt{\langle (\Delta U)^{2} \rangle}},
\end{equation}
and calculated from the canonical ensemble equilibrium fluctuations at $T_{min}$. 
A new class of liquids was recently proposed, namely the class of strongly correlating liquids. These liquids are simple in 
the Roskilde sense of the term\cite{prx} and defined by $R \geq 0.90$. Only 
inverse power-law fluids are perfectly correlating ($R$ = 1), but many models\cite{paper1} as well as experimental liquids\cite{gammagamma} have been shown 
to belong to the class of Roskilde simple liquids. This class is believed 
to include most or all van der Waals and metallic liquids, whereas covalently, hydrogen-bonding or strongly ionic or dipolar liquids are generally 
not Roskilde simple\cite{paper1}. The latter reflects the fact that directional or competing interactions tend to destroy the strong virial/potential energy correlation.  

By plotting $D$ as function of $1-R$, we investigate whether a correlation between the ''simple'' expression of RT for the specific heat and ''Roskilde simple liquids'' exists.
For many of the investigated systems that fall into this simple class such a correlation is \emph{not} expected from the original RT bulk hard-sphere fluid derivation (see below).\newline \newline
\begin{figure}[H]
  \centering
  \includegraphics[width=75mm]{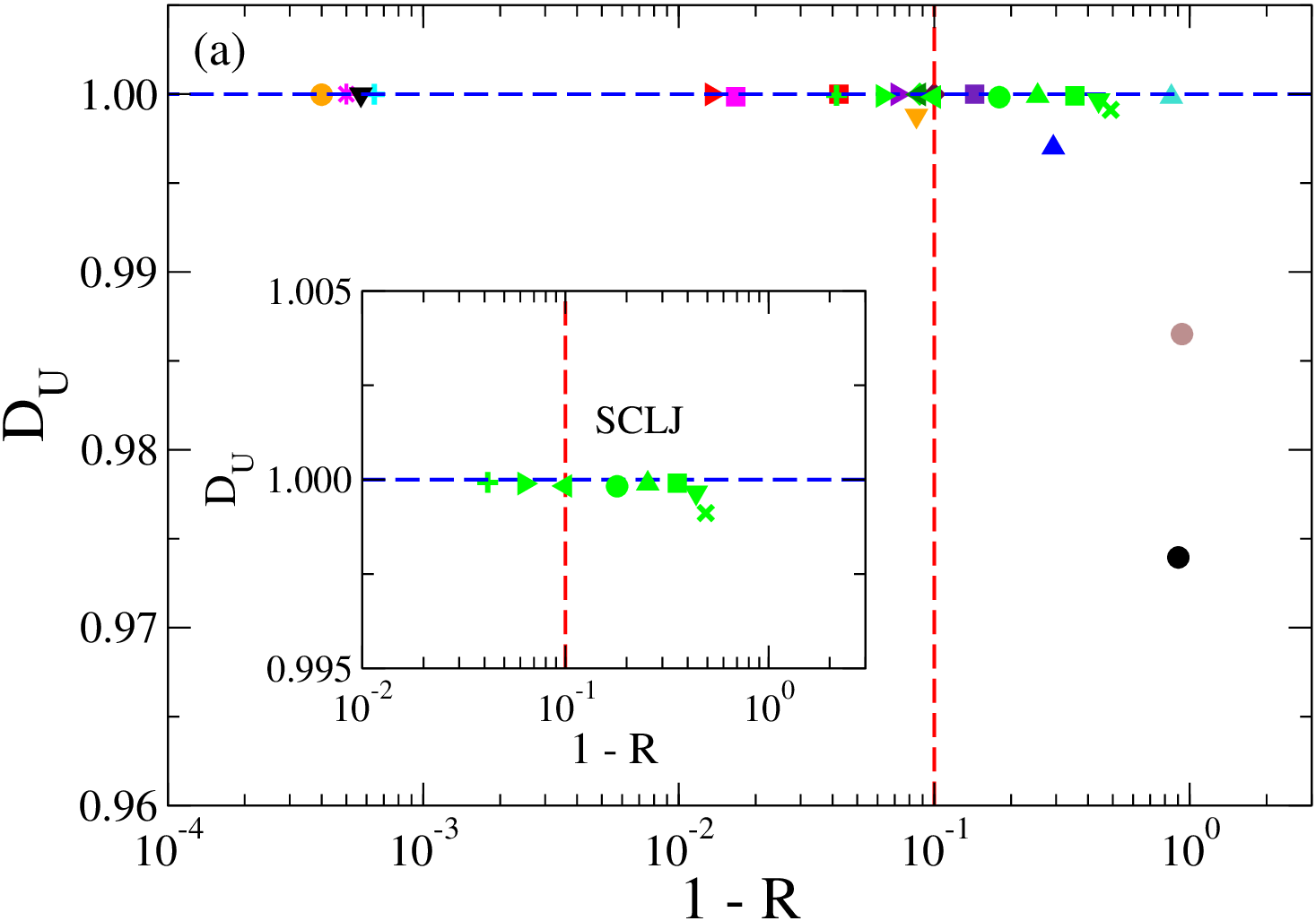}
\end{figure}
\hspace{5pt}
We observe from Fig. \ref{RT2}(a) that for all liquids $D_{U}$  gives a value close to 1, but RT provides
a better approximation for liquids with $R$ larger than $0.90$ (to the left of the red line). A similar behavior 
is observed for $D_{C_{V}^{\mbox{ex}}}$ in Fig. \ref{RT2}(b), however, note the change of scale. The insets of both 
figures show for the SCLJ liquid how RT deteriorates 
as $R$ decreases below 0.90. We conclude that the RT expressions work better for systems that are Roskilde simple at the state points in question. 

Originally\cite{RT}, RT was argued from fundamental-measure thermodynamic 
perturbation theory for bulk hard-sphere fluids and via simulation shown to describe inverse 
power-law systems to a high degree of accuracy. Later investigations showed that RT is a good approximation also for LJ 
liquids. These systems are Roskilde simple, and a recently argued quasi-universality\cite{quasi} for Roskilde simple single-component 
atomic systems implies this behavior. 

We have shown that the key determining factor for  RT is not whether systems are atomic or 
molecular (see results for dumbbell, OTP, and LJC), but rather the degree of strong correlation between 
virial and potential energy. This was shown to be the case even for severely nanoconfined molecular systems which exhibit a completely 
different physics from bulk hard-sphere fluids\cite{ingebrigtsen2013} and are thus \emph{not} expected to satisfy the original RT arguments. The latter is, in particular, true also for 
the elongated non-spherical molecules studied here. The observed correlation between RT and Roskilde simple liquids is thus highly nontrivial.

As a further validation of the above viewpoint, we relate the function $\alpha(\rho)$ in the RT expression 
to $h(\rho)$ for Roskilde simple liquids. For such a liquid, temperature separates\cite{thermoscl} into a product of a function of excess entropy per particle and 
a function of density via $T = f(s_{ex})h(\rho)$. Roskilde simple liquids are characterized by having 
isomorphs to a good approximation\cite{paper4}. Isomorphs are curves in the thermodynamic phase diagram along which structure and 
dynamics in reduced units, as well as some thermodynamic quantities are invariant. Along an isomorph both $C_{V}^{\mbox{ex}}$ and $h(\rho)/T$ are invariant, and consequently one may write
\begin{align}
  C_{V}^{\mbox{ex}} & = F\Big(\frac{h(\rho)}{T}\Big).
\end{align}
Since by the RT expression; $C_{V}^{\mbox{ex}} = 3/5 \alpha(\rho)T^{-2/5} = 3/5\big(\alpha(\rho)^{5/2}/T\big)^{2/5}$, it follows that  $h(\rho) = \alpha(\rho)^{5/2}$ or, equivalently,

\begin{equation}
  \alpha(\rho) = h(\rho)^{2/5}.\label{simRT}
\end{equation}
For a LJ system, it was shown in Refs. \onlinecite{thermoscl} and \onlinecite{beyond} that $h(\rho)$ is given by 

\begin{equation}
  h(\rho) = (\gamma_{0} / 2 - 1)\rho^{4} + (2 - \gamma_{0} / 2)\rho^{2},\label{reph}
\end{equation}
in which $\gamma_{0}$ is calculated from the virial/potential energy fluctuations at $\rho$ = 1 and $T$ = 1 
via $\gamma_{0} = \langle \Delta W \Delta U \rangle / \langle (\Delta U)^{2} \rangle$. 

\begin{figure}[H]
  \centering
  \includegraphics[width=75mm]{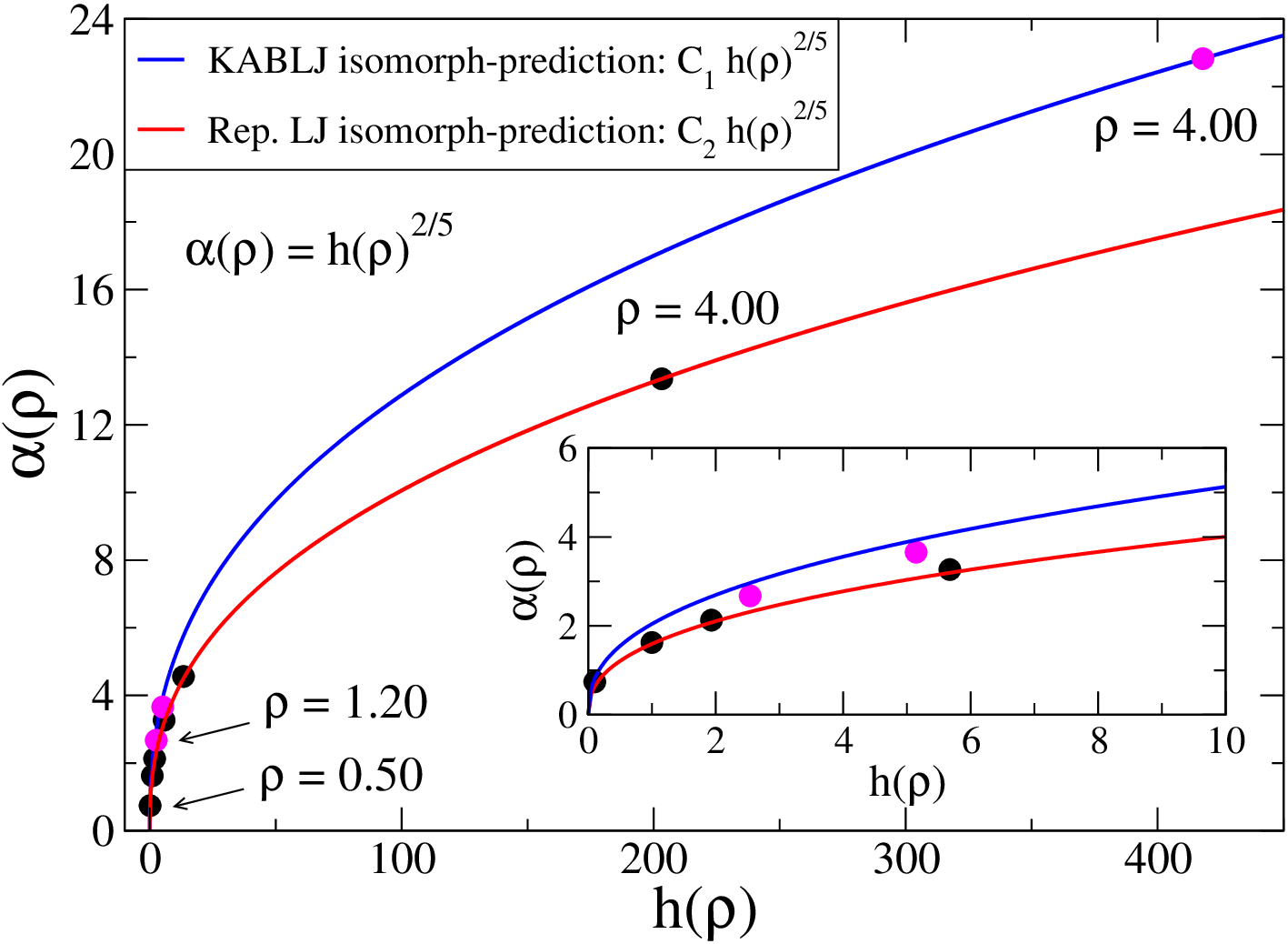}
  \caption{$\alpha(\rho)$ in the RT expression plotted as a function of $h(\rho)$ of the isomorph theory (Eq. (\ref{reph})) for the KABLJ and 
    repulsive LJ system (see text). The red and blue curves are proportional to $h(\rho)^{2/5}$ with the proportionality constant determined from the highest 
    density state point ($\rho$ = 4.00) for each system.}
  \label{RT3}
\end{figure}

Equation (\ref{simRT}) is tested in Fig. \ref{RT3} for the KABLJ and the repulsive LJ system (for which, respectively, $\gamma_{0}$ = 5.35 and $\gamma_{0} = 3.56$). The repulsive LJ system 
is defined from $v(r) = (r^{-12} + r^{-6})/2$ and has $R$ above 99.9\% in its entire phase diagram; $\gamma$ varies from $2$ at low density to $4$ at high density.
We determine $\alpha(\rho)$ for different densities by fitting Eq. (\ref{RT1}) as a function of temperature for each isochore and system. $h(\rho)$ is calculated analytically 
from Eq. (\ref{reph}). Figure \ref{RT3} shows that $\alpha(\rho)$ as predicted by the isomorph theory to a very 
good approximation is given by $h(\rho)^{2/5}$. A complete description is thus given via Eqs. (\ref{RT2}) and (\ref{reph}) for the density and temperature dependence of the 
specific heat, i.e., $C_{V}^{ \mbox{ex}} = (h(\rho)/T)^{2/5}$.

Scaling strategies which relate dynamics to thermodynamics have in the past proven useful to predict perplexing dynamical phenomena. We  
identified here the range of applicability for RT as the class of Roskilde simple liquids. By combining the RT expressions with the isomorph theory, we were able to 
provide also the full density and temperature dependence of the specific heat. The predictive power of the aforementioned scaling strategies 
for most van der Waals and metallic liquids is hereby significantly increased.\newline

\acknowledgments

The center for viscous liquid dynamics ''Glass and Time'' is sponsored by the Danish National Research Foundation via Grant No. DNRF61. We thank Lasse B{\o}hling for 
providing some of the data that establish the background for Fig. \ref{RT3}. Useful discussions with Truls Ingebrigtsen and Jacob Marott are gratefully acknowledged.

%


\end{document}